\documentclass{iopart}

\usepackage{epsfig}

\begin{document}

\title{Recombination Models}

\author{R~J~Fries\dag}

\address{\dag\ School of Physics and Astronomy, University of Minnesota, 
  116 Church Street SE, Minneapolis, MN 55455, USA}

\ead{fries@physics.umn.edu}

\begin{abstract}
We review the current status of recombination and coalescence models that
have been successfully applied to describe hadronization in heavy ion 
collisions at RHIC energies. Basic concepts as well as actual implementations 
of the idea are discussed. We try to evaluate where we stand in our
understanding at the moment and what remains to be done in the future.
\end{abstract}

\submitto{JPG}
\pacs{25.75.Dw,24.85.+p}

\section{Introduction}

Since the last Quark Matter conference, held in 2002,
the idea of quark recombination or coalescence (ReCo)
had a very rapid
and successful career as a model for hadronization. Recombination of quarks
was first formulated over two decades ago by R.\ C.\ Hwa and collaborators 
as a model for hadronization in the fragmentation region of hadron-hadron
collisions \cite{DasHwa:77}. 
There were early efforts to connect it to other descriptions of hadronization 
like fragmentation
\cite{ChaHwa:79,MiJoLa:81}. For heavy ion collisions, quark coalescence ideas 
first appeared soon after that \cite{Gupt:83} and were later on successfully 
used in ALCOR \cite{BiLeZi:95}.

The recent revival of recombination/coalescence began when it was realized
\cite{Voloshin:02} that the elliptic flow $v_2$ measured at RHIC obeys a 
simple valence quark scaling, that naturally arises from a recombination 
picture. Soon after that, it was pointed out \cite{FMNB:03prl,GreKoLe:03prl} 
that other RHIC puzzles, like the anomalous enhancement of baryons and the 
absence of nuclear suppression in baryon spectra can be explained by ReCo 
as well.  

This talk seeks to review the current status of recombination/coalescence 
models. We will revisit the fundamental concepts and experimental findings
that support recombination. We will then discuss different implementations 
of the model and its limitations. We close with
an outlook on future developments.

\section{Basic Ideas}

Hadronization has always been a very difficult aspect of strong interaction
processes. A lot of effort went into the invention of methods to work around
this non-perturbative phenomenon, with the effect that our knowledge of
hadronization dynamics is still very limited.
One example is the formulation of fragmentation or ``quark decay'' functions 
which rely on the common concept of separating long and short distance 
dynamics \cite{CoSo:81}. The cross section of hadron production in 
$e^+ e^-$, lepton-hadron or hadron-hadron collisions can be written as
a convolution
 $ d\sigma_H = d\sigma_a \otimes D_{a\to H}$
of a parton production cross section $d\sigma_a$ with a fragmentation function 
$D_{a\to H}(z,\mu)$ \cite{Owens:86}.
$D_{a\to H}(z,\mu)$ gives the probability to find a hadron $H$ in the 
hadronizing parton $a$ with a momentum fraction $z$, $0<z<1$.
$\mu \gg \Lambda_{\mathrm{QCD}}$ is a perturbative scale that is set by the
transverse momentum.

The $z$-dependence of fragmentation functions is not calculable in 
perturbation theory. However, they are universal, i.e.\ process independent,
objects. Frag\-mentation functions for the most important hadrons have
been measured, mainly in $e^+ e^-$ collisions, and are available in 
parametrized form. Nevertheless, the availability of measured fragmentation 
functions should not lead to the impression that we have a good understanding 
of the underlying hadronization process.

One important question is the following: at which scales (i.e.\ in which 
$P_T$ region) is leading twist perturbative QCD 
(pQCD) reliable?
It has been shown that next-to-leading order (NLO) calculations of 
$\pi^0$ production in $pp$ collisions at RHIC, using modern parametrizations
of fragmentation functions \cite{KKP:00}
are in good agreement with PHENIX data down to surprisingly low pion $P_T$ 
of 1 GeV/$c$ \cite{phenix:03pp}.

On the other hand, there are clear signs that the fragmentation concept is not 
working at very low $P_T$. A good example is the leading particle 
effect. In the very forward (and low $P_T$) region of hadron-hadron
collisions \cite{DasHwa:77}, the composition of particle species deviates from expectations in a fragmentation picture. 
This is impressively confirmed by recent experimental results, e.g.\ from  
the FNAL E791 collaboration \cite{E791:96}. With a 
$\pi^-$ beam impinging on a fixed nuclear target, they measure a $D^-$/$D^+$ 
asymmetry that goes to 1 in the very forward direction. While fragmentation 
would predict this asymmetry to be very close to 0, one can understand this 
effect starting from a recombining $c\bar{c}$ pair produced in the collision.
The recombination $\bar c d \to D^-$ is enhanced with respect to $c \bar d
\to D^+$, because the $d$ is a valence quark in the beam $\pi^-$ while the
$\bar d$ is only a sea quark \cite{BJM:02}. 

Other examples are provided by the RHIC experiments. A proton/pion 
ratio $\sim 1$ was observed in central Au+Au collisions between $P_T=1.5$ and 
4 GeV/$c$ \cite{PHENIX:03prot}. This is in contradiction to pQCD calculations, 
that give $p/\pi \sim 0.1 \ldots 0.2$ \cite{FMNB:03prc}. 
Similar results for $\Lambda/K^0_s$ \cite{STAR:03llbar} suggest that there is 
a general pattern of baryon enhancement at RHIC energies.
This enhancement is so strong that it neutralizes the strong jet quenching
observed for mesons at RHIC \cite{jetqu:03a}. The nuclear modification factor 
$R_{AA}$ for baryons is close to 1 up to 4 GeV/$c$ 
\cite{PHENIX:03prot,STAR:03llbar,STAR:03v2}.
Thus, although pQCD seems to work very well in $p+p$ for $P_T$ above 
1 GeV/$c$, this is not the case in Au+Au even at 4 GeV/$c$.

The two examples have in common that hadronization takes place
in a phase space filled with partons, either beam remnants or the
hot and dense medium created in heavy ion collisions.
This is very different from $e^+ e^-$ collisions where phase space is nearly
empty. One can include corrections to single parton fragmentation in terms of 
higher twist fragmentation. Such contributions include the process
of two or more partons fragmenting into hadrons. However, nothing is known
about these higher twist corrections. 
Instead of the rather complicated twist expansion, let us directly  
look at the limiting case of a phase space densely packed with partons. 
What will happen upon hadronization? 
The most simple picture is that the quarks and antiquarks that 
constitute the valence quark structure of a hadron (having the correct quantum
numbers), recombine/coalesce together to form this hadron. 

\begin{figure}
\begin{center}
  \epsfig{file=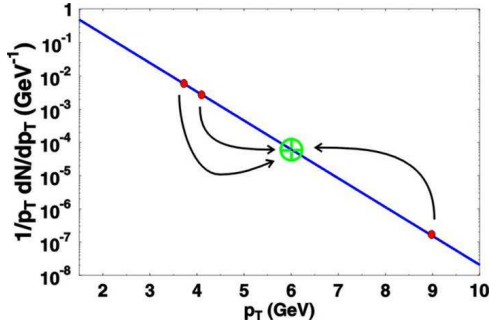,width=6.4cm}
  \caption{\label{fig:1} Recombination and fragmentation for a meson at $P_T 
   \approx 6$ GeV/$c$, starting from a parton spectrum with steep slope 
   (solid line). Fragmentation requires a single parton with transverse 
   momentum larger than $P_T$ to start with, while recombination is possible
   with two partons that have roughly $P_T/2$ each. The competition between
   both processes is decided by slope and normalization of the
   parton spectrum.} 
\end{center}
\end{figure}

In fragmentation the entire parton content of the hadrons has to come from
gluons and $q\bar{q}$ pairs emitted from the fragmenting parton inside the
``black box'' called fragmentation function.
For recombination, it is assumed that no further branching of 
partons occurs. Apparently these are limiting cases. There must be a 
smooth transition between both extremes as a function of phase space density. 
On the other hand, one might 
speculate that the absence of additional branching is a hint that some sort 
of equilibrium is reached in the parton phase.
Figure \ref{fig:1} shows how both processes form a meson.
 

Which partons enter the recombination process? In most 
implementations, quarks are assumed to be non-perturbative and to have an 
effective mass. This, and the fact that only valence quarks are involved, leads
to the interpretation of these degrees of freedom as constituent quarks.
Gluons do not participate in recombination and are used to dress the quarks.
The alternative to the constituent quark picture is to explicitly
convert gluons into $q\bar{q}$ pairs.

\section{Mathematical Formulation}

ReCo can be formulated in terms of Wigner functions. The yield of mesons
$M$ coalescing from two partons $a$, $b$ is given by \cite{FMNB:03prc}
\begin{equation}
  \fl 
  \label{eq:1}
  \frac{dN_M}{d^3P} = \sum_{a,b} \int \frac{d^3 R}{(2\pi)^3} \frac{d^3q d^3r
  }{(2\pi)^3} W_{ab}\left( R-\frac{r}{2},\frac{P}{2}-q;R+\frac{r}{2},\frac{P}{
  2}+q \right) \Phi_M (r,q).
\end{equation}
$W_{ab}$ and $\Phi_M$ are the Wigner functions of the partons and the meson
respectively, $P$ and $R$ are the momentum and spatial coordinate of the 
meson and the sum runs over all possible combinations of quantum numbers,
essentially leading to a degeneracy factor $C_M$.
The generalization of this formula for baryons is straightforward 
\cite{FMNB:03prc}.
The Wigner function for the partons is usually factorized into classical
one-particle phase space distributions, $W_{ab}(r_a,p_a;r_b,p_b)=w_a(r_a,p_a) 
w_b(r_b,p_b)$. This assumes that the partons are completely uncorrelated 
before hadronization. We will come back to this point later.

We can immediately draw some conclusions. Suppose the parton spectrum is
exponential in transverse momentum $p_T$, $w = A e^{-p_T/T}$, with some slope
parameter $1/T$. Then fragmentation and recombination would provide meson 
spectra
\begin{equation}
  \fl
  d N_{\mathrm{frag}}/ d^2 P_T  
   D \propto A e^{-P_T/\langle z\rangle T} \langle
  D\rangle, \qquad   
  d N_{\mathrm{reco}}/ d^2 P_T 
  \propto A^2 e^{-P_T/T}
\end{equation}
respectively. Here $\langle D\rangle$ and $\langle z \rangle$ are average 
values of the fragmentation function and the scaling variable $z$.
Since $\langle z \rangle <1 $, fragmentation is less effective than 
recombination on an exponential spectrum, as long as the normalization $A$ 
is not too small. On the other hand, if the parton spectrum has power law
form, $w=B p_T^{-\alpha}$, the yields are $d N_{\mathrm{frag}}/ d^2 P_T \sim 
P_T^{-\alpha}$ and $d N_{\mathrm{reco}}/ d^2 P_T \sim 
P_T^{-2\alpha}$ for mesons. This implies that fragmentation
will dominate at large $P_T$ for power law spectra (which are predicted by
perturbative QCD). 

If the parton spectrum is exponential, one can check that with increasing
$P_T$ the result is less and less sensitive to the momentum dependence of 
the hadronic Wigner function (i.e.\ the shape of the momentum space 
wave functions used to model it). In fact, it seems that for $P_T > 2$ 
GeV/$c$ one can safely take the momentum space wave function to have zero 
width (i.e.\ to be a $\delta$-function) \cite{FMNB:03prc}.

\begin{figure}
\begin{center}
  \epsfig{file=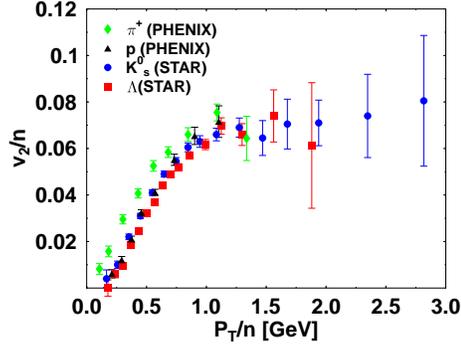,width=6.4cm}
  \caption{\label{fig:2} Elliptic flow $v_2$ scaled with the number $n$ of 
  valence quarks vs $P_T/n$ for four different hadrons.
  Data are taken from STAR ($K^0_s$, $\Lambda+\bar\Lambda$) 
  \cite{STAR:03v2} and PHENIX ($\pi^+$,$p$) \cite{PHENIX:03v2}.} 
\end{center}
\end{figure}

Let us assume that the parton phase exhibits an azimuthal anisotropy. The 
elliptic component of this asymmetry is described by the elliptic
flow coefficient $v_2(p_T)$ \cite{Olli:92}. ReCo predicts the resulting
elliptic flow for hadrons to be 
\cite{Voloshin:02,FMNB:03prc,MoVo:03}
\begin{equation}
  v_2^M (P_T)  = 2 v_2 \left( \frac{P_T}{2} \right) \label{eq:sc1}, \qquad
  v_2^B (P_T)  = 3 v_2 \left( \frac{P_T}{3} \right) \label{eq:sc2}
\end{equation}
for mesons and baryons respectively \cite{FMNB:03prc}.
One should note that the derivation of these scaling laws uses narrow, 
$\delta$-shaped wave functions. See \cite{FMNB:03prc} for the case of
wave functions of finite width.
As can seen be in Figure \ref{fig:2} the scaling laws are impressively 
confirmed by experimental data for $P_T > 1$ GeV/$c$, while at lower $P_T$ the 
mass of the hadron determines the elliptic flow, well described by 
hydrodynamics 
\cite{STAR:03v2,PHENIX:03v2}. Deviations for pions are discussed below.
Above $\sim 5$ GeV/$c$ a perturbative mechanism, driven by jet quenching, 
should take over from recombination, but experimental data are not yet 
conclusive \cite{FMNB:03prc}.

\section{Implementations}

The development of ReCo in the past 18 months was mainly driven by four 
groups who published work using their individual implementations of the 
ReCo concept. These groups are (in parentheses the short form used thereafter)
Duke/Minnesota/Kyoto (Duke) \cite{FMNB:03prl,FMNB:03jpg,FMNB:03prc,
NFB:03,NMABF:03}, Ohio State/Wayne State (Ohio) \cite{MoVo:03,MolLin:03,
Molnar:03},
Oregon \cite{HwaYa:02,HwaYa:03} and Texas A\& M/Budapest (TAM) 
\cite{GreKoLe:03prl,GreKoLe:03prc,GreKoRa:03,GreKo:04}.

The Duke and Oregon groups try to evaluate Equation (\ref{eq:1}) analytically 
using certain assumptions that essentially boil it down to a 
convolution of one-particle distributions and hadron wave functions in 
longitudinal momentum space (longitudinal with respect to the hadron 
momentum). 
As an example, the Duke group writes the meson spectrum
from recombination as \cite{FMNB:03prc}
\begin{equation}
  \fl
  E \frac{dN_M}{d^3P} = C_M \int_\Sigma d \sigma \frac{P\cdot u}{(2\pi)^3)} 
  \int dx \, w_a (\sigma, xP^+) w_b(\sigma,(1-x)P^+) \left| \phi_M(x) \right|^2
\end{equation}
where $d\sigma$ integrates over the hadronization
hypersurface $\Sigma$, $u$ is the four vector orthogonal to $\Sigma$, $x$ 
is the momentum fraction of parton $a$ in the meson, $P^+$ is the light cone 
momentum and $\phi_M$ the wave function of the meson.
The parton phase that undergoes hadronization is assumed to have an
exponential part at low $p_T$ (soft partons) and a power law tail at high 
$p_T$ (hard partons). For central Au+Au collisions at RHIC, the Duke group 
uses a thermal distribution with temperature $T=175$ MeV and average radial 
flow velocity $v=0.55 c$ for the soft partons, while the hard partons are 
taken from a minijet calculation \cite{FMS:02} including energy loss 
\cite{FMNB:03prc}. 

The Ohio and TAM groups developed Monte Carlo implementations of the
recombination process. These can be connected to string or parton cascade 
models that prepare the parton state before hadronization.
One main difference lies in the treatment of the connection between soft and 
hard partons. While
the Duke group strictly separates soft and hard physics, allowing only the
soft partons to recombine and only the hard partons to fragment, the TAM
group includes additional coalescence of soft and hard partons 
\cite{GreKoLe:03prl,GreKoLe:03prc}. The Oregon group carries this a step 
further and replaces fragmentation functions by a scenario where minijet 
partons develop a shower which subsequently recombines. In this model, they 
are able to describe fragmentation functions reasonably well. Applied to 
heavy ion collisions, this allows them to introduce recombination of soft 
partons with shower partons \cite{HwaYa:03}.

\begin{figure}
\begin{center}
  \epsfig{file=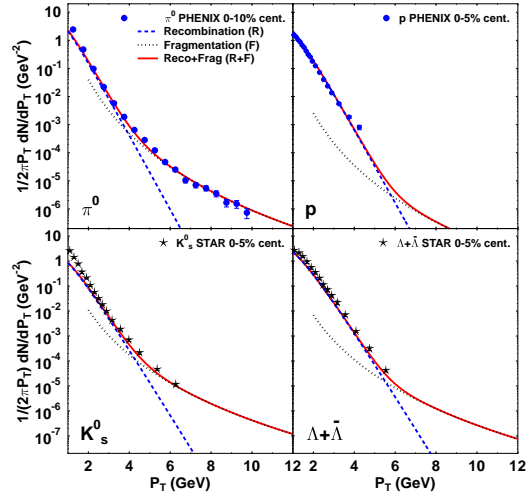,width=7.5cm}
  \caption{\label{fig:3} Spectra of $\pi^0$, $p$, $K_s^0$ and $\Lambda+
  \bar \Lambda$ as a function of $P_T$ at midrapidity for central Au+Au 
  collisions at $\sqrt{s}=200$ GeV/$c$ \cite{FMNB:03prc}. RHIC data are taken
  from PHENIX ($\pi^0,p$) \cite{PHENIX:03pi0,PHENIX:03prot} and
  STAR ($K_s^0$,$\Lambda+\bar \Lambda$) \cite{STAR:03llbar}.} 
\end{center}
\end{figure}

All four groups describe hadron data from RHIC at intermediate and large
transverse momenta very well. The calculations for spectra reproduce the 
noticeable kink in the data around $P_T=4$ GeV/$c$ for mesons and 6 GeV/$c$ for
baryons coming from the transition
from soft (recombination) to hard (perturbative) particle production.
Soft-hard coalescence can improve the fit to data points in
the transition region, which is then extended to even higher $P_T$. 
Figure \ref{fig:3} shows a result obtained by the Duke group compared with
RHIC data. 
The baryon/meson ratios in ReCo are essentially given by the ratio of 
degeneracy factors $C_B/C_M$, leading naturally to an enhanced 
proton/pion ratio. The most recent results can be found in 
\cite{FMNB:03prc,GreKoLe:03prc,HwaYa:03}.

As already mentioned, the elliptic flow of all measured hadron species, 
$\pi$, $K$, $p$, $\Lambda$, $\Xi$ and $\Omega$, 
follows the scaling law (\ref{eq:1}) for $P_T>1$ GeV/$c$. This permits to
unambiguously extract $v_2(p_T)$ for partons. Surprisingly, the elliptic 
flow of strange quarks is the same as for light quarks 
\cite{FMNB:03prc,NFB:03}. Slight deviations from the scaling can be seen
for pions. This can be traced back to the small mass of the pion and the fact
that most pions do not hadronize directly but are from $\rho$ decays
\cite{GreKo:04}.

\section{ReCo Challenges and Answers}

Despite its success, several aspects of ReCo are problematic.
For one, recombination in its current implementation cannot describe
the bulk production of hadronic matter. This seems to 
contradict the claim that ReCo is the right choice for very
dense parton systems. However, one should note that the ReCo 
formula (\ref{eq:1}) is only for the \emph{average} meson created by 
recombination (like the fragmentation formula is for the \emph{average}
hadron produced from one parton). None of these formulas describe the 
\emph{exclusive} hadron content of the system.
Note that in (\ref{eq:1}) the mesons scale with the square of the
parton density, whereas the total number of mesons has to scale linearly, 
of course.
While ReCo as a concept is certainly correct also for the bulk of hadron 
production, the simple formula (\ref{eq:1}) does not describe this.

The second issue is that Equation (\ref{eq:1}) conserves momentum, but not 
energy. In general, it is not possible to have energy conservation in $2\to 1$ 
and $3\to 1$ processes. In reality, particles in this strongly interacting 
environment will be off-shell, making energy conservation possible.
However, instead of taking displacements from the mass shell into account,
it is more convenient to have particles on the mass shell and permit small 
violations of energy conservation. This is a safe procedure as long as the 
violations are small compared to the total energy $E$ of the hadron, i.e.\ 
for $P_T$ above $\sim 2$ GeV/$c$ \cite{FMNB:03prc}. A special role is played 
by Goldstone bosons. Their description is particularly difficult in a picture 
where the coalescing quarks have constituent masses.
 
The situation can be improved by taking into account coalescence of resonances 
and their subsequent decay. This introduces $2\to 2$ and $3\to 2$ processes
and energy can be exactly conserved. Including the $\rho$ resonance, decaying 
into two pions, helps to cure the problem that 
pions are too heavy in a constituent quark picture \cite{GreKo:04}.
Another critical question concerns entropy. Apparently, recombination
reduces the degrees of freedom in the system, therefore leading to a decrease
in entropy. We note that this statement is not relevant as long as we 
do not address total particle numbers. But in 
any case, including resonances considerably improves this situation as well.

\section{For the Future}

One can think of a long list of problems that should be addressed in the
near future. It was pointed out that the study of resonances within the
  recombination model provides new insights about interactions in the
  hadronic phase and could even be used to pin down properties of resonances
  and exotic states \cite{NMABF:03}. 
It also has to be decided whether charmed hadrons follow the recombination
  systematics. The elliptic flow of charm quarks has to be measured 
  \cite{MolLin:03,GreKoRa:03}. First results on higher harmonics are available
now \cite{STAR:03v4}. They provide novel tests for the ReCo concept and can 
be used to further pin down the partonic phase \cite{CheKoLi:03}. 
Furthermore, the role of soft-hard coalescence has to be investigated in 
more detail, in particular with respect to hadron-hadron correlations.

RHIC data indicates that strong hadron-hadron correlations are present at 
intermediate $P_T$, where ReCo dominates \cite{STAR:02cor}. Such correlations 
are expected for the fragmentation process where several hadrons emerge from 
the same parton. It was argued that hadrons from recombination
are emitted statistically so that no correlations should be observable. 
This is not true. Correlations in the parton phase are naturally translated
into correlations in the hadron phase by the recombination process.
We immediately conclude that there must be non-negligible correlations
between soft partons.

Let us assume we want to describe the recombination of four partons 1,2,3,4
into two mesons $A$, $B$. In analogy to Equation (\ref{eq:1}), 
a 4-parton Wigner function $W_{1234}$ is needed to describe double meson 
production. We can include correlations in the parton phase by replacing
the simple single particle factorization by
\begin{equation}
  W_{1234} = N \prod_{i=1}^4 w_i \left( 1+ \sum_{i<j} C_{ij}\right)
\end{equation}
where $C_{ij}$ is a 2-parton correlation function and $N$ is a normalization 
factor. This will lead to non trivial correlation functions between
mesons. The correlations in the parton phase could originate from interactions
between hard and soft partons and would be naturally present in soft-hard
recombination.
It remains to be seen whether the correlations measured at RHIC
will consistently fit into this picture. But if this is the case, ReCo
will provide a fascinating new picture of the partonic phase.

What will happen at LHC? Part of the ReCo success story is, that jet quenching
is so strong at RHIC, suppressing hard processes in the final state by a 
factor of 4. Hard processes will be more abundant at LHC, but 
increased jet quenching and a brighter thermal source will probably 
overwhelm them up to even higher $P_T$ than at RHIC \cite{FM:03lhc}

\section{Summary}

Quark recombination/coalescence is a successful model to describe hadronization
in dense parton systems.
Central Au+Au collisions provide a partonic medium that is sufficiently
dense for coalescence of soft partons to overcome fragmentation at 
intermediate transverse momenta between 2 and 
5 GeV/$c$. Soft-hard coalescence could push this region to even 
higher $P_T$. 

\ack
I am indebted to my collaborators B.\ M\"uller, 
S.\ A.\ Bass, C.\ Nonaka, and M.\ Asakawa. I would like to thank all 
participants of the workshop {\it Quark Recombination} held at the Institute 
for Nuclear Theory, University of Washington, December 8--9 2003 for 
stimulating discussions. This work was supported by DOE grant 
DE-FG02-87ER40328.

\end{document}